\documentclass[superscriptaddress,aps,pra,twocolumn,showpacs,floatfix]{revtex4-2}
\usepackage[utf8]{inputenc}
\usepackage{graphicx,amsmath,amsfonts,amssymb}
\usepackage{color}
\usepackage[colorlinks=true, allcolors={blue}]{hyperref}
\usepackage{graphicx}
\usepackage{epstopdf}
\usepackage{float}
\usepackage{placeins}
\usepackage[english]{babel}

\usepackage{lipsum}
\DeclareMathOperator{\sech}{sech}

\newcommand{\orcid}[1]{\href{https://orcid.org/#1}{\includegraphics[width=7pt]{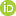}}}

\begin{document}

\preprint{APS/123-QED}

\title{Towards gravimetry enhancement with squeezed states}

\author{Oziel R. de Araujo\orcid{0000-0002-5475-9878}}
\affiliation{Departamento de F\'{i}sica, Universidade Federal do Piau\'{i}, Campus Ministro Petr\^{o}nio Portela, 64049-550, Teresina, PI, Brazil}
% \affiliation{Instituto Estadual de Educação, Ciência e Tecnologia do Maranhão - IEMA, Avenida Castelo Branco, 65430-000, Vargem Grande, MA, Brasil}

\author{Lucas S. Marinho\orcid{0000-0002-2923-587X}}
%\email{lucas.marinho@ufpi.edu.br}
\affiliation{Departamento de F\'{i}sica, Universidade Federal do Piau\'{i}, Campus Ministro Petr\^{o}nio Portela, 64049-550, Teresina, PI, Brazil}

\author{Jonas F. G. Santos\orcid{0000-0001-9377-6526}}
\affiliation{Faculdade de Ci\^{e}ncias Exatas e Tecnologia, Universidade Federal da Grande Dourados, Caixa Postal 364, 79804-970, Dourados, MS, Brazil}

\author{Carlos H. S. Vieira\orcid{0000-0001-7809-6215}}
\email{carloshsv09@gmail.com}
%\email{vieira.carlos@ufabc.edu.br}
\affiliation{Centro de Ci\^{e}ncias Naturais e Humanas, Universidade Federal do ABC,
Avenida dos Estados 5001, 09210-580 Santo Andr\'e, S\~{a}o Paulo, Brazil}
\affiliation{Department of Physics, State Key Laboratory of Quantum Functional Materials,
and Guangdong Basic Research Center of Excellence for Quantum Science,
Southern University of Science and Technology, Shenzhen 518055, China}


\begin{abstract}
We investigate the sensitivity of gravitational acceleration estimation using squeezed probe states in a quantum metrology framework. In particular, we analyze how the squeezing phase, beyond its amplitude, affects the attainable precision. We show that probes squeezed along the canonical phase-space quadratures can surpass the shot-noise limit only in specific time regimes, whereas position–momentum correlated input states can consistently overcome this limit across all interaction times. Furthermore, we demonstrate that optimal sensitivity can be achieved by combining projective momentum measurements with a time-dependent adjustment of the squeezing phase. Our results highlight the fundamental role of phase-engineered squeezing in quantum gravimetry protocols and provide new insights into the design of optimized sensing strategies.
\end{abstract}

                              %display ded
\maketitle

%\tableofcontents

\textit{Introduction.}~The pursuit of novel protocols towards high-precision measurements of gravitational acceleration is crucial in the field of gravimetry~\cite{Markus_Advac202,Wei_PhysRevR25,Joachim_arxiv25,Cassens_PhysRevX25}. Experimentally, significant advances have been made in gravimeters employing several quantum techniques, including atomic interferometry~\cite{APeters_2001,Biedermann_PRA15,Bidel_AppliedPhy13}, Bose–Einstein condensate (BEC)~\cite{Cassens_PhysRevX25,Anders_PRL21,Abend_PRL16}, ultracold neutrons~\cite{Ichikawa_prl14},   superconducting systems~\cite{Merlet2021}, and optomechanics systems~ \cite{Qvarfort2018,Cervantes_AppliedPhys14}. Theoretically, improvements in the sensing of gravitational fields are pivotal in fundamental tests of general relativity theory~\cite{LIGO,LIGO2023,Ciufolini2016,Dimopoulos_PRL07,Dimopoulos_PRD08}, ultimate sensing precision in quantum metrology~\cite{VMontenegro_PRR25}, and also in the ongoing challenges to unify quantum mechanics and general relativity~\cite{Biswas_PRL12,Woodard_2009,Carlip_2008,Vlatko_RevMPhy25}. Furthermore, from a technological perspective, ultrasensitive gravimeters are essential for developments in geological surveys and navigation, for instance~\cite{Crowley_Geophysical06,Tapley_science06,Corentin_Geophysics}.

Accurate estimation of physical quantities is crucial for theoretical developments and experimental breakthroughs. In estimation theory, the Cramér-Rao inequality, $\text{Var}( \hat{\alpha})\ge [n\mathcal{I}_{\alpha}(x)]^{-1/2}$, establishes an essential lower bound on the sensitivity (i.e., the standard deviation) with which an unknown parameter, $\alpha$, encoded in a conditional probability $p(x|\alpha)$ of observing the random value $x$, can be estimated~\cite{Helstrom1969,Holevo_book}. Here, $\hat{\alpha}$ denotes an unbiased estimator that maps the outcome $x$ to an estimate of the parameter $\alpha$, $n$ is the number of independent experiment trials, and $\mathcal{I}_{\alpha}(x)$ is the classical Fisher information (CFI). The CFI quantifies how changes in the parameter value $\alpha$ affect the probability distribution of the measured quantity $x$, through a particular choice of measurement strategy or a set of positive operator-valued measurements (POVM)~\cite{Seth_PRL06,Giovannetti2011}. Given its dependence on the specific measurement strategy, the CFI can be optimized over all possible POVM~\cite{Helstrom1969,Seth_PRL06,Giovannetti2011}, leading to an ultimate upper limit for the CFI, namely Quantum Fisher information (QFI), $ \mathcal{F}_\alpha =  \text{Tr}[\hat{\rho}_\alpha \hat{L}_\alpha^2 ]$,
where $\hat{\rho}_\alpha$ is the density operator encoded with the parameter $\alpha$ and $\hat{L}_\alpha$ is the symmetric logarithmic derivative defined as $\partial_\alpha \hat{\rho}_\alpha = ( \hat{L}_\alpha \hat{\rho}_\alpha + \hat{\rho}_\alpha \hat{L}_\alpha)/2$ \cite{Milburn1996,Seth_PRL06,Giovannetti2011}.
The quantum parameter estimation theory establishes the QFI as a fundamental precision limit for estimating the parameter $\alpha$, bounding the classical Fisher information, i.e., $\mathcal{F}_\alpha \geq \mathcal{I}_\alpha$ and giving rise to the quantum Cramér–Rao inequality, which defines the minimal achievable estimation variance employing a quantum state~\cite{Braunstein_CavesPRL1994,Seth_PRL06,Giovannetti2011,Milburn1996,jiang2014quantum}. 

Quantum metrology aims to explore the features of quantum probes such as entanglement~\cite{Pezze_PRL09,Seth_PRL06,Giovannetti2011, Chattopadhyay2025-hf}, coherence~\cite{Carlos_PRA2025}, non-classicality~\cite{Braun_RMPhy18,Pezze_RMPhy18, xdnc-tc2y}, contextuality~\cite{Jae2024}, and squeezing~\cite{Caves1981,Wineland_PRA97,MA201189,Zhou_PRL25}, to overcome the precision of sensing unknown parameters compared to their classical counterparts~\cite{jiang2014quantum,Seth_PRL06,Giovannetti2011}, leading to a series of experimental advances in this field~\cite{Cappellaro_RMPhys17,Milburn_book}. In particular, since their original conception~\cite{Slusher_PRL85,Shelby_PRL86} and pioneering proposal for use in laser interferometric gravitational wave detectors~\cite{Caves1981}, squeezed states have found applications in a wide range of gravimetry protocols~\cite{Schnabel2010}. Despite the recognized advantage of employing squeezed probe states in gravimetry protocols, most applications have focused mainly on the squeezing amplitude or on situations where the squeezing phase is aligned with a canonical phase-space quadrature, such as position or momentum. 

In this Letter, we investigate the boost in the estimation of the gravitational acceleration, $g$, by considering input squeezed states endowed with a squeezing phase that differs from the canonical quadratures of position and momentum. Specifically, we analyze how the gravitational acceleration is dynamically encoded in both the amplitude and phase of the squeezed probe and how different phases of squeeze affect the QFI. We demonstrate that by choosing an appropriate phase of the squeezed input state, one can create a non-zero correlation between the position and momentum quadratures, thereby significantly increasing the QFI and improving the sensitivity in estimating the gravitational acceleration, thereby overcoming vacuum probe states, regardless of the free-falling time under the gravitational field. Additionally, we show that a projective momentum measurement can saturate the QFI bound and therefore achieve this optimal sensitivity. Notably, standard quadrature squeezing along position or momentum does not always yield optimal results, highlighting the importance of the fine-tuning of the squeezing phase in quantum metrology protocols.

\textit{The model.}~We consider the dynamics of a quantum particle with inertial mass, $m$, free falling in a non-relativistic gravitational potential, $U(z)=mgz$, where $z$ is the distance from the floor and $g$ denotes the gravitational acceleration. Experimentally, this scenario can be feasible by employing ultra-cold atoms that freely fall from a vertical tower in the gravitational potential~\cite{Abend_PRL16,Ichikawa_prl14}. The system is initialized in a Gaussian state, 
\begin{equation}
\rho_{0}(z_{0},z_{0}^{\prime})=\mathcal{N}_{0}\exp\left[-\frac{(z_{0}^{2}+z_{0}^{\prime2})}{4\sigma^{2}}+\frac{i\gamma(z_{0}^{2}-z_{0}^{\prime2})}{4\sigma^{2}}\right],
\label{GSS}
\end{equation}
that formally corresponds to a squeezed vacuum state characterized by the initial position uncertainty $\sigma=\sigma_{0}[\cosh(2r)-\sinh(2r)\cos(2\theta)]^{1/2}$ and the position-momentum correlation factor $\gamma=\sinh(2r)\sin(2\theta)$~\cite{DODONOV1980150,Dodonov2,Dodonov2014,Dodonov_2002,supp_material,AraujoMPLA2019,MarinhoPRA2020,LustosaPRA2020,PP2024EPJP,Marinho2024SciRep}. Here, $\mathcal{N}_{0}=1/(\sqrt{2\pi}\sigma)$ denotes the normalization constant, $r$ and $\theta$ represent the squeezing amplitude and phase, respectively, and $\sigma_{0}$ is the standard deviation of the vacuum state. The phase $\theta$ sets the quadrature along which squeezing is applied and determines the orientation of the uncertainty ellipse in phase space~(see Supplemental Material (SM)~\cite{supp_material}). For instance, when $\theta =0$ the squeezing is along the position quadrature, $\sigma=\sigma_{0}/e^{2r}$ (reduction in position uncertainty), and for $\theta=\pi/2$, the squeezing is along the momentum quadrature, $\sigma=\sigma_{0}e^{2r}$ (increasing in position uncertainty). For intermediate values of $\theta$, the squeezed input state (\ref{GSS}) exhibits a non-null correlation between the position and momentum quadratures~\cite{Bohm_book}. Recent studies have highlighted the role of such position-momentum correlations as a valuable resource in metrological tasks~\cite{Porto2025Scripta,ThiagoNJP2025,PortoPRA2025,porto2025multiparameterestimationpositionmomentumcorrelated}. Position-momentum correlations are a key resource for universal continuous-variable quantum computation (CVQC)~\cite{upreti2025interplayresourcesuniversalcontinuousvariable}, as established by a recent resource theory~\cite{upreti2025symplecticcoherencemeasurepositionmomentum}.

\begin{figure*}[!ht]
\centering
\includegraphics[width=1.0\linewidth]{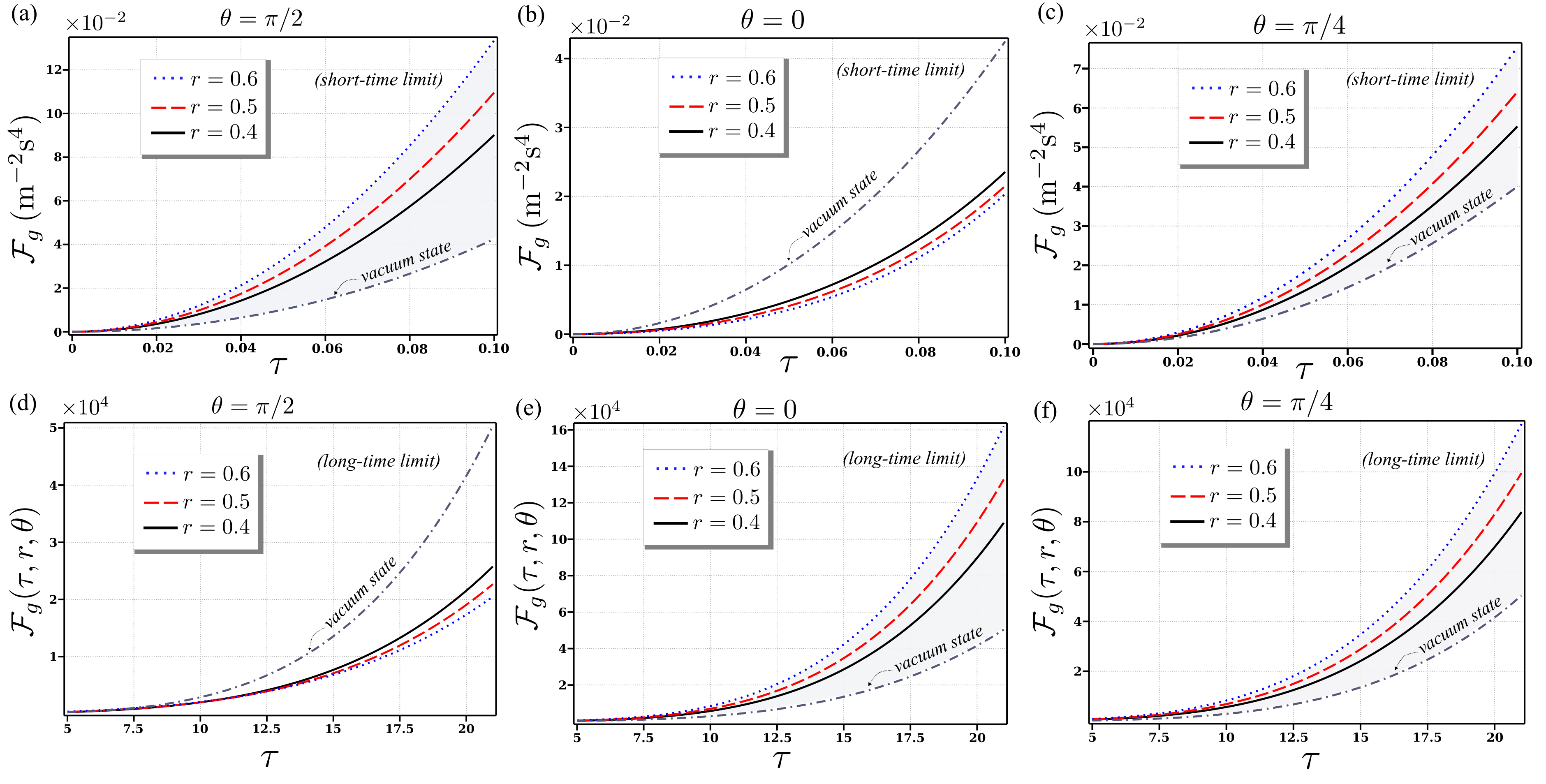}
\caption{QFI for estimation of $g$ as a function of the evolution time $\tau/\tau_{0}$ for different amplitudes and phases of the squeezed input state. Panels (a-c) correspond to the short-time evolution regime, while panels (d-f) depict the dynamics at the long-time limit. Curves represent squeezing intensities $r=0.4$ (solid black line), $r=0.5$ (dashed red line), $r=0.6$ (dotted blue line), for distinct squeezing phases $\theta$ of the squeezed input state~(\ref{GSS}) and vacuum state $r=0$ (dot-dashed gray line). Both panels illustrate that the quantum advantage in estimating gravitational acceleration with squeezed probe states initialized in canonical directions, such as position ($\theta = 0$) and momentum ($\theta=\pi/2$), is limited to a certain time [shaded area in Fig.~\ref{fig01}(a, e)]. In contrast, probes with position-momentum correlation offer a consistent improvement in the QFI over any vacuum state, regardless of the evolution time~[shaded area in Fig.~\ref{fig01}(c, f)].}
\label{fig01}
\end{figure*}

\textit{Gravitational acceleration estimation.}~Using the path integral formalism, the evolution of the input squeezed state~(\ref{GSS}) under a uniform gravitational field can be analytically determined~(see SM for more details~\cite{supp_material}). Additionally, the QFI concerning the parameter $g$ is given by~(details in SM~\cite{supp_material})
\begin{equation}
\begin{split}
\mathcal{F}_{g}(\tau,r,\theta) &=
\frac{\sinh^{2}(2r)\sin^{2}(2\theta)+1}{4\sigma_{0}^{2}[\cosh(2r)-\sinh(2r)\cos(2\theta)]}\tau^{4} \\
&\quad
+\frac{2\tau_0\sinh(2r)\sin(2\theta)}{\sigma_{0}^2}\tau^{3}\\
&\quad
+\frac{4\tau_{0}^{2}[\cosh(2r)+\sinh(2r)\cos(2\theta)]}{\sigma_{0}^{2}}\tau^{2},
\label{QFIg}
\end{split}
\end{equation}
where $\tau$ is the evolution time and $\tau_{0}=m\sigma_{0}^2/\hbar$ is a constant with time dimension. In the particular case where $r=0$ (i.e., $\gamma=0$), the QFI~(\ref{QFIg}) becomes $\mathcal{F}_{g}^{\text{vac}}(\tau)=\tau^4/4\sigma^2_0+4\tau_0^2\tau^2/\sigma^2_0$, consistent with the result obtained in Ref.~\cite{Paris2025arxiv} for an input vacuum state. Some conclusions can be drawn from Eq.~(\ref{QFIg}). First, the QFI depends on both amplitude and phase of the squeezing. Consequently, it allows setting different squeezing phases for the input probe, rather than restricting squeezing to the conventional quadratures by fixing $\theta=0$ or $\pi/2$, thereby enhancing the sensitivity to gravitational acceleration relative to the input vacuum state. Moreover, since the energy of the input probe state is independent of $\theta$, possible advantages in the QFI by changing the squeezing phase come without any additional energy cost. Second, the QFI is insensitive to the intensity of the gravitational acceleration, meaning that the capacity to estimate the $g$ does not change with the strength of this parameter. Third, for short ($\tau\ll\tau_{0}$) and long ($\tau\gg\tau_{0}$) interaction times limit, the QFI~(\ref{QFIg}) evolves into $\mathcal{F}_{g}^{\tau\ll\tau_{0}}(\tau,r,\theta)\approx[\sigma^{2}/\sigma_{0}^{2}]\mathcal{F}_{g,\tau\ll\tau_{0}}^{\text{vac}}(\tau)$ and $\mathcal{F}_{g}^{\tau\gg\tau_{0}}(\tau,r,\theta)\approx[\sigma_{0}^{2}/\sigma^{2}+\gamma^{2}]\mathcal{F}_{g,\tau\gg\tau_{0}}^{\text{vac}}(\tau)$, respectively. This characteristic enables fine-tuning of the squeezing intensity $r$ and phase $\theta$ of the input probe to enhance beyond the vacuum state, thereby increasing the sensitivity to $g$ at specific interaction times. 

Figure~\ref{fig01} presents the QFI~(\ref{QFIg}) for estimating the gravitational acceleration parameter $g$ as a function of the dimensionless interaction time $\tau/\tau_{0}$, considering different input probe states. In all plots, we set the following parameters $m=1$, $\sigma_0=1$, $\hbar=1$ to focus only on the general trends and features of the dynamics. Panels (a-c) correspond to the dynamic of $\mathcal{F}_{g}(\tau,r,\theta)$ for short interaction times, whereas panels (d-f) illustrate the behavior for long times. The curves represent the squeezing intensities $r=0.4$ (solid black line), $r=0.5$ (dashed red line), $r=0.6$ (dotted blue line), for distinct squeezing phases of the input state~(\ref{GSS}), and the vacuum state $r=0$ (dot-dashed gray line). For the input states squeezed in momentum ($\theta=\pi/2$), the QFI surpasses that of the vacuum probe only in the short-time regime, $\mathcal{F}_{g}^{\tau\ll\tau_{0}}(\tau,r,\pi/2)\approx e^{2r}\mathcal{F}_{g,\tau\ll\tau_{0}}^{\text{vac}}(\tau)$ [shaded area in Fig.~\ref{fig01}(a)], the input vacuum state being more sensitive in estimating the gravitational acceleration $g$ for long times, $\mathcal{F}_{g}^{\tau\gg\tau_{0}}(\tau,r,\pi/2)\approx e^{-2r}\mathcal{F}_{g,\tau\gg\tau_{0}}^{\text{vac}}(\tau)$ [Fig.~\ref{fig01}(d)]. Conversely, for probes squeezed in position ($\theta=0$), the QFI enhancement over the input vacuum state occurs exclusively in the long time limit, $\mathcal{F}_{g}^{\tau\gg\tau_{0}}(\tau,r,0)\approx e^{2r}\mathcal{F}_{g,\tau\gg\tau_{0}}^{\text{vac}}(\tau)$ [shaded area in Fig.~\ref{fig01}(e)], with the input vacuum state more sensitive for short times, $\mathcal{F}_{g}^{\tau\ll\tau_{0}}(\tau,r,0)\approx e^{-2r}\mathcal{F}_{g,\tau\ll\tau_{0}}^{\text{vac}}(\tau)$ [Fig.~\ref{fig01}(b)]. Interestingly, probe states initially endowed with position-momentum correlations ($\theta=\pi/4$) achieve a QFI greater than that of the vacuum state in both the short, $\mathcal{F}_{g}^{\tau\ll\tau_{0}}(\tau,r,\pi/4)\approx\cosh(2r)\mathcal{F}_{g,\tau\ll\tau_{0}}^{\text{vac}}(\tau)$ [Fig.~\ref{fig01}(c)], and long, $\mathcal{F}_{g}^{\tau\gg\tau_{0}}(\tau,r,\pi/4)\approx(\sinh^{2}2r+\sech 2r)\mathcal{F}_{g,\tau\gg\tau_{0}}^{\text{vac}}(\tau)$ [Fig.~\ref{fig01}(f)], interaction-time regimes, respectively. This result highlights that the quantum advantage in estimating gravitational acceleration using squeezed probes initialized in the canonical position ($\theta=0$) and momentum ($\theta=\pi/2$) directions is limited to a specific time window. In contrast, squeezed probes endowed with initial position-momentum correlation can achieve a consistent improvement in the QFI over the vacuum state, regardless of the interaction time. 

Importantly, the QFI does not necessarily increase monotonically with the squeezing strength $r$. This behavior arises from the nontrivial interplay between the hyperbolic functions of $r$ and the trigonometric dependence on the squeezing phase $\theta$ in Eq.~\ref{QFIg}, which contribute with different weights depending on the interaction time. As a result, increasing $r$ may either enhance or reduce the sensitivity, depending on the squeezing orientation and the time regime. This explains why, in some cases, smaller values of $r$ may outperform larger ones, as observed in Figs. \ref{fig01}(b) and \ref{fig01}(d). This non-monotonic behavior further underscores the central role of the squeezing phase in optimizing metrological performance.

\textit{Relative quantum Fisher information.} To quantify the possible enhancement from squeezed states in estimating the gravitational acceleration beyond the vacuum probe, we define the relative QFI~(RQFI)
\begin{equation}
\mathcal{Q}(\tau,r,\theta)=\frac{\mathcal{F}_{g}(\tau,r,\theta)}{\mathcal{F}_{g}^{\text{vac}}(\tau)}.
\label{RQFI}
\end{equation}
For $\mathcal{Q}>1$, the squeezed probe provides a precision advantage over the vacuum state for estimating gravitational acceleration. The case $\mathcal{Q}<1$ represents no advantage of employing squeezed states, while $\mathcal{Q}=1$ corresponds to the equivalence between both the vacuum and the squeezed probe in the estimation of $g$. 

\begin{figure}[!h]
\centering
\includegraphics[width=0.8\linewidth]{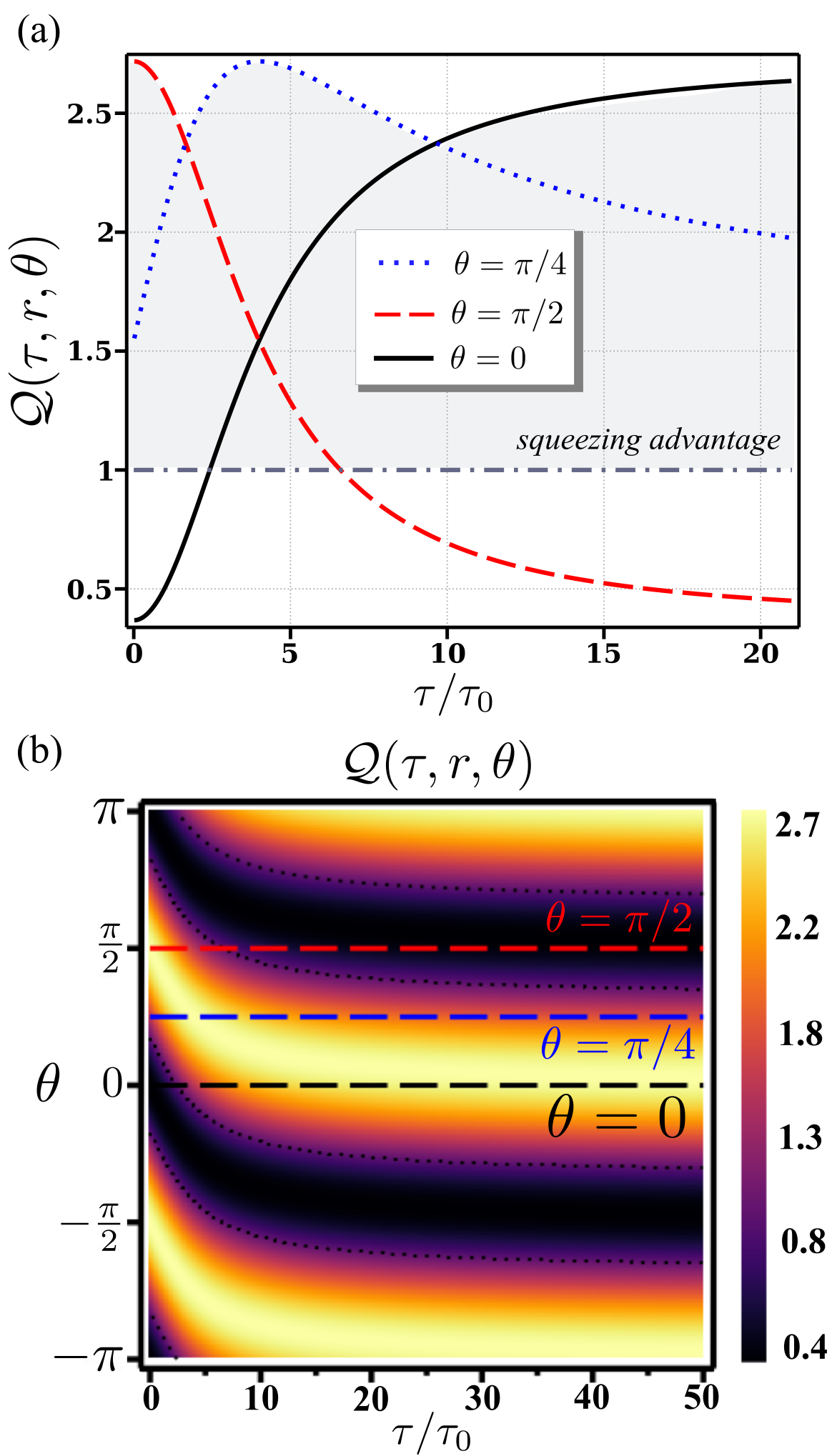}
\caption{The Relative Quantum Fisher Information (RQFI) $\mathcal{Q}$, with values $\mathcal{Q} > 1$ implies an advantage using squeezed states, $\mathcal{Q} < 1$ indicates no gain, and $\mathcal{Q} = 1$ denotes equal performance with vacuum states in the estimation of $g$. Panel (a) illustrates the temporal evolution of $\mathcal{Q}$ for different squeezing directions in phase space. The solid black line corresponds to squeezing applied along the canonical position quadrature direction ($\theta = 0$), the dashed red line represents squeezing along the momentum quadrature ($\theta = \pi/2$), and the dotted blue line depicts a probe exhibiting position-momentum correlations, with the squeezing direction chosen at an intermediate phase of $\theta = \pi/4$. In (b), we display a contour plot of $\mathcal{Q}$ as a function of time and squeezing phase $\theta$. It highlights that multiple directions beyond $\theta = \pi/4$ yield $\mathcal{Q} > 1$, demonstrating the broad flexibility in optimizing squeezing orientation for enhanced metrological performance. In all plots, we fixed the squeezing amplitude at $r=0.5$.}
\label{fig02}
\end{figure}

Figure~\ref{fig02}(a) illustrates the time evolution of $\mathcal{Q}(\tau, r, \theta)$ for three distinct squeezing phases, where the behaviors in short ($\tau\ll \tau_{0}$) and asymptotic regimes ($\tau\gg \tau_{0}$) are summarized in Table~\ref{tab:QFI_leading}. We consider probe states with squeezing applied along the position quadrature (solid black line, $\theta = 0$), the momentum quadrature (dashed red line, $\theta = \pi/2$), and an intermediate value corresponding to the position-momentum correlated state (dotted blue line, $\theta = \pi/4$). These results highlight the fact that the sensitivity of the gravitational acceleration strongly depends on the phase of the input squeezed probe. For a squeezed probe in position ($\theta=0$), the RQFI monotonically increases with time, reaching its asymptotic limit $e^{2r}$, whereas for a probe squeezed in momentum $\theta=\pi/2$ it decays toward $e^{-2r}$, indicating a loss of metrological advantage $Q<1$. As a result, the squeezing advantage $\mathcal{Q}>1$ (shaded area) provided by probe states with squeezing in position or momentum crucially depends on the interaction time. Conversely, by employing the position-momentum correlated state ($\theta=\pi/4$), with the same quantity of squeezing, it always exhibits a squeezing advantage to estimate the gravitational acceleration $g$ in relation to the vacuum probe, regardless of the evolution time. 
\begin{table}[t]
\caption{$\mathcal{Q}(\tau,r,\theta)$ for different squeezing phases in the short and asymptotic time regimes.}
\label{tab:QFI_leading}
\begin{ruledtabular}
\begin{tabular}{ccc}
$\theta$ & short-time ($\tau \ll \tau_0$) & asymptotic ($\tau \gg \tau_0$) \\
\colrule
$0$ & $e^{-2r}$ & $e^{2r}$ \\[4pt]
$\pi/2$ & $e^{2r}$ & $e^{-2r}$ \\[4pt]
$\pi/4$ & $\cosh(2r)$ & $\cosh^{2}(2r)$ \\
\end{tabular}
\end{ruledtabular}
\end{table}

Despite our choice of $\theta=\pi/4$ in Fig.~\ref{fig02}(a), there are other position-momentum correlated states with a QFI surpassing that of the vacuum state. Figure~\ref{fig02}(b) displays a contour plot of $\mathcal{Q}$ as a function of both time $\tau$ and the squeezing phase $\theta$, allowing us to explore the continuous variation of metrological performance in different squeezing orientations. In particular, the results reveal that the enhanced sensitivity for all times ($\mathcal{Q} > 1$) is not always limited to the specific case $\theta = \pi/4$, but can be achieved across a wide range of angles, highlights that optimal metrological gain for estimation of gravitational acceleration does not depend solely on the squeezing strength but critically on the phase-space orientation of the squeezed quadratures.

\textit{Measurement schemes.}~Although the QFI establishes the ultimate precision bound for an unbiased estimator, $\hat{\alpha}$, the optimal measurement required to achieve this bound is often difficult or impractical~\cite{Seth_PRL06,Giovannetti_LloydPRL2006,PaternostroARXIV2025}. This is because it requires a projective measurement associated with POVM elements given by projectors onto the eigenstates of the logarithmic derivative operator $\hat{L}_{\alpha}$. To identify measurement schemes capable of approaching the bound imposed by the QFI, we focus on strategies that can be readily described within the Gaussian formalism \cite{PaternostroPRA2017}. We consider local Gaussian projective measurements, which can be efficiently modeled by updating the covariance matrix as $\widetilde{\boldsymbol{\sigma}} = \boldsymbol{\sigma} + \boldsymbol{\sigma}_m$ \cite{PaternostroPRA2017,PaternostroARXIV2025}, where the matrix $\boldsymbol{\sigma}_m = \text{Diag}[s, s^{-1}]/2$ characterizes the measurement and is parametrized by $s$. This framework encompasses projective position measurements ($s \rightarrow 0$), projective momentum measurements ($s \rightarrow \infty$), as well as homodyne and heterodyne detection ($s = 1$). The exact CFI, $\mathcal{I}_{g}(\tau,r,\theta,s)$, that quantifies the information extracted by a given measurement strategy \cite{PaternostroPRA2017,PaternostroARXIV2025} can be analytically determined in terms of the parameter $s$ (see SM~\cite{supp_material}). By scanning the values of $s$, it is possible to identify which measurement strategy is most effective in achieving the ultimate bound, where the CFI approaches the QFI. 
\begin{figure*}[!ht]
\centering
\includegraphics[width=1.0\linewidth]{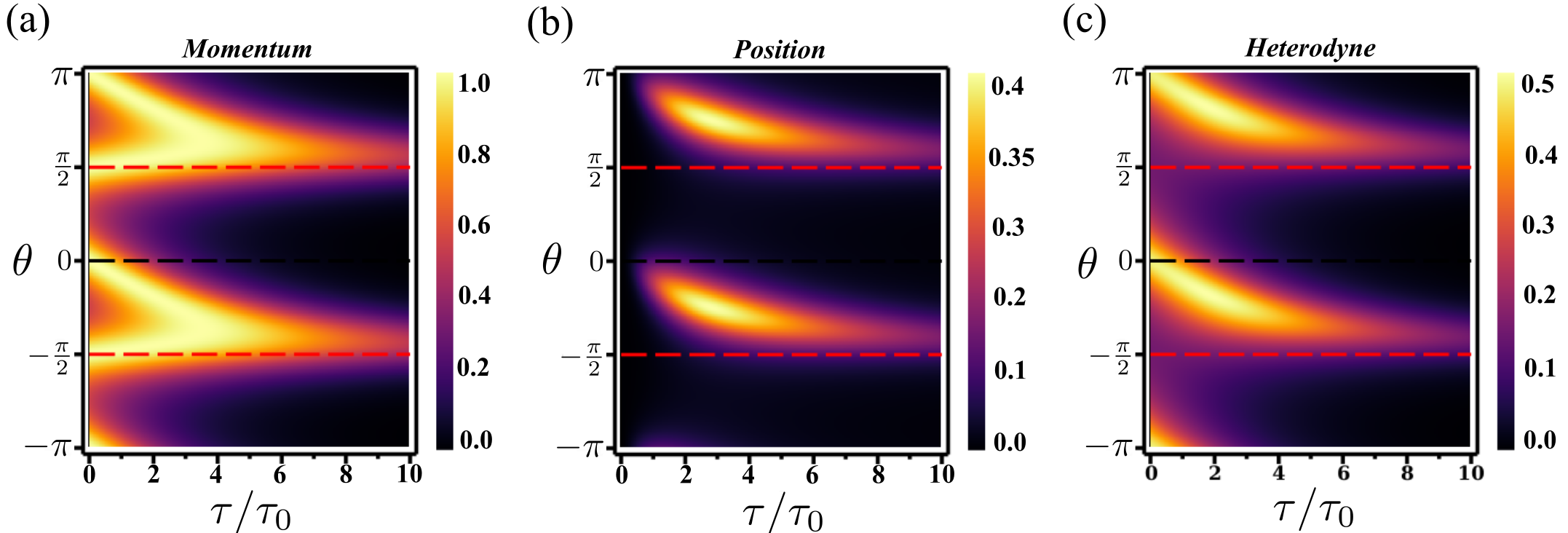}
\caption{Behavior of the ratio $\mathcal{R}(\tau,r,\theta,s) = \mathcal{I}_{g}(\tau,r,\theta,s)/\mathcal{F}_g(\tau,r,\theta)$ between the CFI and the QFI bound for different measurement schemes, as a function of the squeezing phase $\theta$ and the evolution time $\tau$. Here, the squeezing parameter $r=0.5$ is kept fixed across all panels, while the measurement setting $s$ varies from panel to panel. Panel (a) shows the case $s \rightarrow \infty$ (projective momentum measurement), (b) corresponds to $s \rightarrow 0$ (projective position measurement), and (c) illustrates the heterodyne scheme with $s = 1$. Note that only the momentum measurement strategy can saturate the QFI bound, yielding the maximum possible information [$\mathcal{R} = 1 $, yellow region in Fig.~\ref{fig3}(a)]. Yet, the optimal choice of squeezing phase is time-dependent, and squeezing along canonical quadratures, position ($\theta=0$) or momentum ($\theta=\pi/2$), does not always provide the best sensitivity at all times.}
\label{fig3}
\end{figure*}

In Fig.~\ref{fig3}, we show the ratio $\mathcal{R}(\tau,r,\theta,s) = \mathcal{I}_{g}(\tau,r,\theta,s)/\mathcal{F}_g(\tau,r,\theta)$ between the CFI, $\mathcal{I}_{g}(\tau, r, \theta, s)$, and the QFI bound, $\mathcal{F}_{g}(\tau, r, \theta)$, for different measurement strategies. The ratio $\mathcal{R}$ quantifies how closely a given measurement approaches the ultimate precision bound set by the QFI.  Here, the squeezing parameter $r=0.5$ is kept fixed across all panels, while the measurement setting $s$ varies between them. We analyze how different squeezing phases $\theta$ of the squeezed input state can be adjusted over time to reach the QFI limit. The limiting situation $s \rightarrow \infty$, which denotes an ideal momentum measurement, is shown in Fig.~\ref{fig3}(a).  This approach can theoretically reach the highest sensitivity allowed by quantum physics in this domain, since the CFI can saturate the QFI bound, i.e., $\mathcal{I}_{g} = \mathcal{F}_{g}$ ($\mathcal{R} = 1$) [lighter yellow region, in Fig.~\ref{fig3}(a)].  However, this optimality depends on the appropriate choice of the squeezing phase, which changes with time $\tau$.  The complementary scenario $s\rightarrow 0$, corresponding to a projective position measurement, is displayed in Fig.~\ref{fig3}(b).  Unlike the momentum example, the projective position measurement does not saturate the QFI constraint, $\mathcal{R}\approx0.4$ [yellow region in Fig.~\ref{fig3}(b)]. The heterodyne detection scheme with $s = 1$ is shown in Figure~\ref{fig3}(c).  This approach also fails to reach the QFI limit [$\mathcal{R}\approx0.5$, yellow region in Fig.~\ref{fig3}(c)], highlighting the importance of searching for appropriate measurement strategies in quantum metrology. Interestingly, optimal sensitivity is not always achieved by canonical quadrature squeezing, i.e., squeezed probe with phase along position ($\theta = 0$) or momentum ($\theta = \pi/2$).  In addition, the time dependence of the optimal phase $\theta_\text{opt}(\tau)$ suggests that adaptive or dynamically controlled squeezed strategies may be needed to approach the quantum limit over different time scales consistently.

From an experimental perspective, the protocol discussed here is compatible with current quantum gravimetry platforms, such as atom interferometers and atomic fountains~\cite{APeters_2001,Biedermann_PRA15,Bidel_AppliedPhy13,Abend_PRL16}, as well as other freely falling ultracold atomic systems~\cite{Ichikawa_prl14}~(see SM~\cite{supp_material} for a simulation considering an ensemble of cesium atoms). In these setups, motional Gaussian states can be engineered and manipulated with high precision, including the generation of squeezed states and controlled phase-space rotations~\cite{Caves1981,Wineland_PRA97,MA201189,Schnabel2010}. Moreover, projective momentum measurements can be implemented using standard time-of-flight or Doppler-sensitive detection schemes, as routinely employed in ultracold atom experiments~\cite{APeters_2001,Bloch2008}, while Gaussian measurement strategies, such as homodyne and heterodyne detection, are well established in continuous-variable quantum optics~\cite{Lvovsky2009,Weedbrook}. These considerations indicate that the phase-engineered squeezing strategies proposed here are, in principle, experimentally accessible.

\textit{Conclusion.}~In this work, we have employed squeezed probe states to estimate the gravitational acceleration parameter. We analyze how both the amplitude and phase of the squeezing affect the estimation sensitivity. Our results show that the metrological advantage in gravitational acceleration estimation, compared to a vacuum probe state, is not solely determined by the squeezing strength but crucially depends on the squeezing phase, i.e., the direction of the applied squeezing in the probe state. In particular, while probes squeezed along the canonical quadratures may fail to surpass the shot-noise limit, position–momentum correlated states with the same squeezing magnitude can exhibit a quantum Fisher information exceeding that of the vacuum state. Moreover, we identified optimal measurement strategies and squeezing phases that allow the classical Fisher information to reach the quantum limit. These findings deepen our understanding of how the phase of squeezed probes can be exploited to enhance precision in quantum gravimetry protocols. In realistic implementations, environmental noise and decoherence degrade both the squeezing strength and phase-space correlations, gradually reducing the QFI and eventually suppressing the quantum advantage. Although our analysis focuses on the ideal unitary regime, our results provide a fundamental benchmark for quantum-enhanced gravimetry and suggest that optimizing the squeezing phase may help mitigate decoherence-induced losses in practical scenarios.

\textit{Acknowledgments.}
O.R.A. acknowledges the Graduate Program in Physics at Federal University of Piau\'i. L.S.M. acknowledges support from the National Institute of Science and Technology on the National Institute of Photonics (INFO) CNPq - INCT grant 409174/2024-6. J. F. G. S. acknowledges CNPq Grant No. 420549/2023-4, Fundect Grant No. 83/026.973/2024, and Universidade Federal da Grande Dourados for support.
C.H.S.V. acknowledges the São Paulo Research Foundation (FAPESP) Grant No. 2023/13362-0 and Grant No. 2025/14546-2 for financial support and the Southern University of Science and Technology (SUSTech) for providing the workspace during the internship.

%\newpage
\bibliography{references}% Produces the bibliography via BibTeX.

% %\begin{widetext}
% \newpage
%\appendix
% \twocolumngrid

%%%%%%%%% Merge with supplemental materials %%%%%%%%%%
\newpage
\pagebreak
%\widetext
\onecolumngrid
%%%%%%%%%% Merge with supplemental materials %%%%%%%%%%

 \newpage
 \begin{center}
 \vskip0.5cm
 {\Large Supplemental Material: Towards gravimetry enhancement with squeezed states}
 \vskip0.2cm
 Oziel. R. de Ara\'ujo,$^{1,2}$ Lucas S. Marinho,$^1$ Jonas F. G. Santos,$^3$ Carlos H. S. Vieira$^{4,5}$
 \vskip0.1cm
 \textit{$^1$ Departamento de F\'{i}sica, Universidade Federal do Piau\'{i}, Campus Ministro Petr\^{o}nio Portela, \\ 64049-550, Teresina, PI, Brazil,\\ $^2$ Instituto Estadual de Educação, Ciência e Tecnologia do Maranhão - IEMA, Avenida Castelo Branco, \\ 65430-000, Vargem Grande, MA, Brasil\\ $^3$ Faculdade de Ci\^{e}ncias Exatas e Tecnologia, Universidade Federal da Grande Dourados, Caixa Postal 364, \\ 79804-970, Dourados, MS, Brazil,\\ $^4$ Centro de Ci\^{e}ncias Naturais e Humanas, Universidade Federal do ABC,
 Avenida dos Estados 5001, \\ 09210-580 Santo Andr\'e, S\~{a}o Paulo, Brazil,\\ $^{5}$ Department of Physics, State Key Laboratory of Quantum Functional Materials,
 and Guangdong Basic Research Center of Excellence for Quantum Science,
 Southern University of Science and Technology, \\ Shenzhen 518055, China}
 \vskip0.1cm
 \end{center}
 \vskip0.4cm

 %%%%%%%%%% Prefix a "S" to all equations, figures, tables and reset the counter %%%%%%%%%%
 \setcounter{equation}{0}
 \setcounter{figure}{0}
 \setcounter{table}{0}
 \setcounter{page}{1}
 \renewcommand{\theequation}{S\arabic{equation}}
 \renewcommand{\thefigure}{S\arabic{figure}}

 In this Supplementary Material, we provide additional information and calculations supporting the statements made in this Letter. In Section I, we describe our model, the time evolution through a gravitational potential using Feynman's propagator, and the calculation of the covariance matrix used to characterize the Gaussian squeezed state. Section II presents the main results on quantum parameter estimation, introduces the Quantum Fisher Information (QFI), and provides the general formulas employed throughout the manuscript. Finally, in Section III, we discuss several measurement schemes that could achieve the ultimate precision allowed by quantum mechanics.

\section{The model, time evolution, and covariance matrix}

\begin{figure}[!h]
\centering
\includegraphics[width=1.0\linewidth]{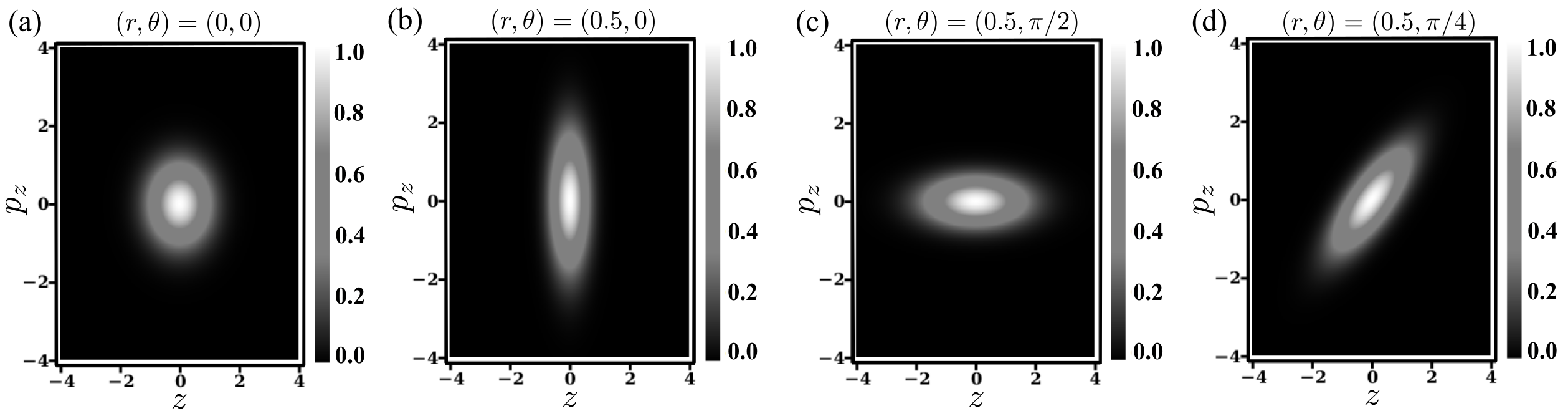}
\caption{Wigner function of the squeezed input state~(\ref{eq01_SM}) for different values of amplitude $r$ and phase $\theta$, showing that, for example, in the specific case $\theta = \pi/4$, the state exhibits position-momentum correlations, resulting in an elliptical shape with the quadratures correlated along this direction in phase space. }
\label{fig01SM}
\end{figure}

Here, we describe the time evolution of a quantum particle with inertial mass $m$, free-falling in a non-relativistic gravitational potential, $U(z)=mgz$, where $g$ denotes the gravitational acceleration and $z$ is the distance from the floor. This potential represents an approximation of Newton's gravitational potential, $U(z)=-GMm/(R+z)$, where $M$, $R$, and $G$ are the mass and radius of the Earth, and $G$ is Newton’s constant, respectively. The system has Hamiltonian $H=\hat{p}_z^2/2m + mg\hat{z}$ and is initialized in the following Gaussian state, 
\begin{equation}
\rho_{0}(z_{0},z_{0}^{\prime})=\mathcal{N}_{0}\exp\left[-\frac{(z_{0}^{2}+z_{0}^{\prime2})}{4\sigma^{2}}+\frac{i\gamma(z_{0}^{2}-z_{0}^{\prime2})}{4\sigma^{2}}\right],
\label{eq01_SM}
\end{equation}
that corresponds to a squeezed vacuum state in the coordinate representation. The parameters $r$ and $\theta$ represent the squeezing amplitude and phase, respectively, $\mathcal{N}_{0}=1/\sqrt{2\pi}\sigma$ denotes the normalization constant, $\sigma=\sigma_{0}[\cosh(2r)-\sinh(2r)\cos(2\theta)]^{1/2}$ is the position uncertainty at initial time, $\sigma_0$ the standard deviation of the vacuum state, and $\gamma=\sinh(2r)\sin(2\theta)$ denotes the position-momentum correlation parameter~\cite{DODONOV1980150,Bohm_book}. For the initial state (\ref{eq01_SM}), position-momentum correlation is encoded by $\sigma_{zp_z}=\langle\hat{z}\hat{p}_{z}+\hat{p}_{z}\hat{z}\rangle/2-\langle\hat{z}\rangle\langle\hat{p}_{z}\rangle=\hbar\gamma$, and this state reaches the minimum of the Schrödinger-Robertson uncertainty relation, $\sigma_z\sigma_{p_z}-\sigma_{zp_z}^2\ge \hbar^2/4$, where $\sigma_z$ and $\sigma_{p_z}$ are variances of coordinate and momentum, respectively~\cite{DODONOV1980150,Dodonov2,Dodonov2014,Dodonov_2002}. In turn, in the special case of $\gamma=0$, the initial state~(\ref{eq01_SM}) turns into an uncorrelated Gaussian state. Physically, the phase of the squeezing $\theta$ sets the quadrature along which the squeezing is applied and determines the orientation of the uncertainty area in phase space. Figure~\ref{fig01SM} illustrates the Wigner function of the initial state~(\ref{eq01_SM}) for different values of the squeezing amplitude and phase. Figure~\ref{fig01SM}(a) depicts the case where $(r,\theta)=(0,0)$, corresponding to the vacuum state. Figure~\ref{fig01SM}(b) illustrates the squeezing along the position quadrature (reduction in position uncertainty) while Fig.~\ref{fig01SM}(c) displays the squeezing along the momentum quadrature (increase in position uncertainty). Consequently, when the ellipse axes are aligned to the canonical quadratures in phase space, the correlation coefficient equals zero. For intermediate values of $\theta$, such as $\theta=\pi/4$, the squeezed input state (\ref{eq01_SM}) exhibits non-zero correlations between position and momentum quadratures, represented by the uncertainty ellipse non-parallel to the coordinate and momentum quadrature in phase space, Fig.~\ref{fig01SM}(d).

In the Feynman path integral framework, the propagator $K(z,t;z_0,t_0)$ corresponds to the probability amplitude for a particle to travel from a spatial point $z_0$ at time $t_0$ to another point $z$ at a later time $t$~\cite{Feynman_book}. Defining the evolution time as $\tau = t - t_0$, the propagator that describes the time evolution of a particle with mass $m$ in a uniform gravitational field $g$ is given by~\cite{Feynman_book}
\begin{equation}
K(z,t;z_0',t_0) =
\sqrt{\frac{m}{2\pi i\hbar\tau}}
\exp\bigg\{
\frac{i}{\hbar}\bigg[
\frac{m(z-z_{0}')^{2}}{2\tau}
-\frac{mg\tau(z+z_{0}')}{2}
-\frac{mg^{2}\tau^{3}}{24}
\bigg]\bigg\},
\end{equation}
and the time-evolved state yields
\begin{equation}
\rho(z,z^{\prime},\tau)=\int K(z,z_{0},t)\rho_{0}(z_{0},z_{0}^{\prime})K^{\ast}(z^{\prime},z_{0}^{\prime},t_{0})dz_0dz^{\prime}_0.
\end{equation}
Considering the squeezed probe state (\ref{eq01_SM}), the evolved state becomes 
\begin{equation}
\rho_{g}(z,z^{\prime},\tau)=\mathcal{N}_{\tau}\exp\left[-\mathcal{A}_{1}\left(z+\frac{\mathcal{A}_{2}}{2\mathcal{A}_{1}}\right)^{2}\right]
\exp\left[-\mathcal{A}_{1}^{\dagger}\left(z^{\prime}+\frac{\mathcal{A}_{2}^{\dagger}}{2\mathcal{A}_{1}^{\dagger}}\right)^{2}\right],
\label{final_state}
\end{equation}
where
\begin{equation}
\mathcal{N}_{\tau}=\left(\mathcal{N}_{0}e^{\mathcal{A}_{3}-\frac{\mathcal{A}_{2}^{2}}{4\mathcal{A}_{1}}}\right)\left(\mathcal{N}_{0}e^{\mathcal{A}_{3}-\frac{\mathcal{A}_{2}^{2}}{4\mathcal{A}_{1}}}\right)^{\dagger}, \;\;\;\;\;\;\;\;  \mathcal{N}_{0}=(\sqrt{2\pi}\sigma)^{-1},
\end{equation}

\begin{equation}
\mathcal{A}_{1}=a_{1}-ib_{1},\;\;\; \mathcal{A}_{1}=a_{2}+ib_{2},\;\;\;\mathcal{A}_{3}=a_{3}+ib_{3},
\end{equation}
\begin{equation}
a_{1}=-\frac{m^{2}\sigma^{2}}{2\tau^{2}\hslash^{2}(\gamma^{2}+1)+4\hslash\gamma m\sigma^{2}\tau+2m^{2}\sigma^{4}},\;\;\;a_{2} = g\tau^2a_{1},\;\;\;a_{3} = \frac{g^{2}\tau^{4}}{4}a_{1}.
\end{equation}

\begin{equation}
b_{1} = \frac{m\tau\hslash(\gamma^{2}+1)+m^{2}\gamma\sigma^{2}}{2\tau^{2}\hslash^{2}(\gamma^{2}+1)+4\hslash\gamma m\sigma^{2}\tau+2m^{2}\sigma^{4}},
\end{equation}
\begin{equation}
b_{2} = -\frac{g\left(\tau^{3}m\hslash^{2}(\gamma^{2}+1)+3\hslash\gamma m^{2}\sigma^{2}\tau^{2}+2m^{3}\tau\sigma^{4}\right)}{2\tau^{2}\hslash^{3}(\gamma^{2}+1)+4\hslash^{2}\gamma m\sigma^{2}\tau+2\hbar m^{2}\sigma^{4}},
\end{equation}
\begin{equation}
b_{3} = -\frac{g^{2}\left(-\tau^{5}m\hslash^{2}(\gamma^{2}+1)+\hslash\gamma m^{2}\sigma^{2}\tau^{4}+2m^{3}\tau^{3}\sigma^{4}\right)}{12\tau^{2}\hslash^{3}(\gamma^{2}+1)+24\hslash^{2}\gamma m\sigma^{2}\tau+12m^{2}\hbar\sigma^{4}}.
\end{equation}

Using the evolved state (\ref{final_state}), the mean vector $\boldsymbol{d}_{\tau}=(\langle x\rangle,\langle p\rangle)^T$ and the covariance matrix elements $\boldsymbol{\sigma}_{\tau}=\langle d_{i}d_{j}+d_{j}d_{i}\rangle-2\langle d_{i}\rangle\langle d_{j}\rangle$ at time $\tau$ are given by
\begin{equation}
 \boldsymbol{d}_{\tau}(\tau,r,\theta) = (-g\tau^2/2, -mg\tau)^{T}   
\end{equation}
and
\begin{equation}
\sigma_{11}(\tau,r,\theta)=\frac{\sigma^{2}(\gamma^{2}+1)}{2}\frac{\tau^{2}}{\tau_{0}^{2}}+2\sigma^{2}\gamma\frac{\tau}{\tau_{0}}+2\sigma^{2}, \;\;\;\;\;\;\;\;  \sigma_{22}(\tau,r,\theta)=\frac{(\gamma^{2}+1)}{2}\frac{m^{2}\sigma^{2}}{\tau_{0}^{2}},
\end{equation}
\begin{equation}
\sigma_{12}(\tau,r,\theta)=\sigma_{21}(\tau,r,\theta)=\frac{\hbar(\gamma^{2}+1)}{2}\frac{\tau}{\tau_{0}}+\gamma\hbar,
\end{equation}
%\begin{equation}
%\sigma_{22}(\tau,r,\theta)=\frac{(\gamma^{2}+1)}{2}\frac{m^{2}\sigma^{2}}{\tau_{0}^{2}},
%\end{equation}
where $\sigma=\sigma_{0}[\cosh(2r)-\sinh(2r)\cos(2\theta)]^{1/2}$ and $\gamma = \sinh(2r)\sin(2\theta)$.

\section{Quantum parameter estimation and Gaussian States}

This section provides a brief overview of tools for quantum parameter estimation using Gaussian states. Due to its paramount significance, the search for strategies to enhance measurement precision has been an ongoing endeavor in physics. In a classical parameter estimation theory, Fisher information (FI) measures the amount of information about an unknown parameter obtained through a particular measurement strategy~\cite{Helstrom1969,Holevo_book}. Let $p(x|\alpha) = \text{Tr}[\hat{\rho}(\alpha) \hat{\Pi}_x]$ denote the conditional probability of obtaining the measurement outcome $x$ given a particular value of the parameter $\alpha$, and $\hat{\rho}(\alpha)$ be the density operator that describes the
state of the system. In quantum theory, the measurement process is described using a set of positive operator-valued measures (POVMs) ${\hat{\Pi}_x}$~\cite{Nielsen_Chuang_book}, parameterized by $x$. These operators are positive semi-definite and satisfy the completeness relation $\int dx \;  \hat{\Pi}_x = \hat{I}$, which ensures normalization. The Classical Fisher Information (CFI), which quantifies the sensitivity of the probability distribution with respect to changes in $\alpha$, is defined as \cite{Holevo_book}
\begin{gather}
\mathcal{I}_{\alpha} = \int dx \;  \frac{1}{p(x|\alpha)} \left( \frac{\partial p(x|\alpha)}{\partial \alpha} \right)^2. \label{eq:CFI}
\end{gather}
Concerning the estimation of the parameter $\alpha$, the associated data $x$ are useful when $p(x|\alpha)$ exhibits a noticeable peak in reaction to changes in $\alpha$. In contrast, let $p(x_i|\alpha)$ be roughly flat. Then, to estimate the value of $\alpha$, which can only be precisely ascertained by averaging throughout the entire set of samples, numerous realizations of $x$ are needed.
In this framework, the Cramér-Rao inequality \cite{cramer1946} provides a lower bound on the standard deviation $\Delta \alpha = \sqrt{\langle \alpha^2 \rangle - \langle \alpha \rangle^2}$ of any unbiased estimator of $\hat{\alpha}$:
\begin{equation}\label{EQ_CRI}
\Delta \alpha \geq \frac{1}{\sqrt{n \mathcal{I}_{\alpha}}},
\end{equation}
where $n$ is the number of independent measurements in the experiment. This inequality demonstrates that Fisher information is a crucial figure of merit in these situations, since it establishes a fundamental bound on the possible precision in parameter estimation. The Cramér-Rao constraint only applies to unbiased estimators, i.e., estimators that satisfy $\langle \alpha_{est} \rangle = \alpha_{real}$.

Quantum Fisher Information (QFI) represents the maximum amount of information on an unknown parameter $\alpha$ that can be assessed using a quantum state~\cite{Braunstein_CavesPRL1994,Giovannetti_LloydPRL2006}. The QFI, denoted as $\mathcal{F}_{\alpha}$, is defined by taking the maximum of the Classical Fisher Information (CFI) in Eq.(\ref{eq:CFI}) over all possible sets of POVMs ${\hat{\Pi}_x}$\cite{Helstrom1969,Holevo_book,Braunstein_CavesPRL1994,Milburn1996}:

\begin{equation}
\mathcal{F}_\alpha = \max_{\{\hat{\Pi}_x\}} \;\mathcal{I}_\alpha =  \text{Tr}[\hat{\rho}_\alpha \hat{L}_\alpha ],
\end{equation}
where the symmetric logarithmic derivative $\hat{L}_\alpha$ is defined
by $\partial_\alpha \hat{\rho}_\alpha = ( \hat{L}_\alpha \hat{\rho}_\alpha + \hat{\rho}_\alpha \hat{L}_\alpha)/2$.
This formulation establishes the QFI as a measure of the ultimate precision achievable in estimating the parameter $\alpha$ within quantum mechanics. Since the QFI results from an optimization over all allowable quantum measurements, it serves as an upper bound to the classical Fisher information, i.e., $\mathcal{F}_\alpha \geq \mathcal{I}_\alpha$~\cite{Braunstein_CavesPRL1994}. It is important to note that the QFI provides only an optimal bound~\cite{Pezze_RMPhy18}, although it does not specify which measurement strategy achieves it, a task the CFI addresses. This results from the fact that the ideal measurement corresponds to a projective measurement defined by POVM elements given by the projectors onto the eigenstates of the symmetric logarithmic derivative (SLD) operator $L_\alpha$, which, however, may not be directly implementable in practice~\cite{PaternostroARXIV2025}. In the particular case, the QFI for an unknown parameter $\alpha$ in a pure Gaussian state can be written as~\cite{jiang2014quantum,Carlos_PRA2025,monras2013phase,Porto2025Scripta}
\begin{equation}
\mathcal{F}_{\alpha}(\boldsymbol{\sigma},\mathbf{d})=\frac{\text{Tr}[(\boldsymbol{\sigma}^{-1}\partial_{\alpha}\boldsymbol{\sigma})^{2}]}{4}+2(\partial_{\alpha}\boldsymbol{d})^{\text{T}}(\boldsymbol{\sigma^{-1}})(\partial_{\alpha}\boldsymbol{d}),
\label{QFIgaus}
\end{equation}
where $\boldsymbol{d}=(\langle \hat{q}\rangle, \langle \hat{p}\rangle)^{T}$ is the mean vector,  $\boldsymbol{\sigma}_{ij}=\langle \hat{d}_{i}\hat{d}_{j}+\hat{d}_{j}\hat{d}_{i}\rangle-2\langle \hat{d}_{i}\rangle\langle \hat{d}_{j}\rangle$ is the covariance matrix, where $\hat{d}_i$ simply denotes the quadrature operators with $\hat{d}_1=\hat{q}$ and $\hat{d}_2=\hat{p}$, and $\partial_{\alpha}\equiv\partial/\partial_{\alpha}$ represents the partial derivative with respect to $\alpha$, respectively.

Considering the estimation of the gravitational acceleration $g$ for the evolved squeezed vacuum state and its first, $\boldsymbol{d}_{\tau}(\tau,r,\theta)$, and covariance matrix, $\boldsymbol{\sigma}_{\tau}(\tau,r,\theta)$, the QFI (\ref{QFIgaus}) becomes
\begin{equation}
\mathcal{F}_{g}(\tau,r,\theta) =
\frac{\sinh^{2}(2r)\sin^{2}(2\theta)+1}{4\sigma_{0}^{2}[\cosh(2r)-\sinh(2r)\cos(2\theta)]}\tau^{4} 
+\frac{2m\sinh(2r)\sin(2\theta)}{\hbar}\tau^{3}
+\frac{4[\cosh(2r)+\sinh(2r)\cos(2\theta)]\tau_{0}^{2}}{\sigma_{0}^{2}}\tau^{2},
\end{equation}
where $r$ and $\theta$ represent the squeezing amplitude and phase, respectively. One of the main objectives of this work is to investigate not only the squeezing amplitude but also the squeezing phase, which can enhance the quantum Fisher information (QFI), as most existing studies focus primarily on the squeezing amplitude.

By taking the limit 
$r=0$ and $\theta=0$, we obtain the corresponding quantum Fisher information (QFI) for the vacuum state as follows
\begin{equation}
\mathcal{F}_{g}^{\text{vac}}(\tau)=\frac{\tau^{4}}{4\sigma_{0}^{2}}+\frac{4m^{2}\sigma_{0}^{2}\tau^{2}}{\hbar^{2}}.
\end{equation}
This quantity provides the shot-noise limit, which is the fundamental sensitivity bound that can be reached with uncorrelated particles and classical (i.e., coherent) states of light. In the short-time regime ($\tau \rightarrow 0$) and the long-time regime ($\tau \rightarrow \infty$), the QFI can be expressed in terms of the corresponding value for the vacuum state as
\begin{gather}
\mathcal{F}_{g}^{\tau\rightarrow0}(r,\theta,\tau)\approx\mathcal{F}_{g,\tau\rightarrow0}^{\text{vac}}(\tau)\frac{\sigma^{2}}{\sigma_{0}^{2}} \;\;\;\;\;\;\;\text{and} \;\;\;\;\;\;\;\; \mathcal{F}_{g}^{\tau\rightarrow\infty}(r,\theta,\tau)\approx\mathcal{F}_{g,\tau\rightarrow0}^{\text{vac}}(\tau)\left(\frac{\sigma_{0}^{2}}{\sigma^{2}}+\gamma^{2}\right).
\end{gather}

These expressions are useful as they clearly reveal the conditions under which the shot-noise limit can be surpassed by adjusting the squeezing amplitude and/or phase.

\begin{figure}[!h]
\centering
\includegraphics[width=0.5\linewidth]{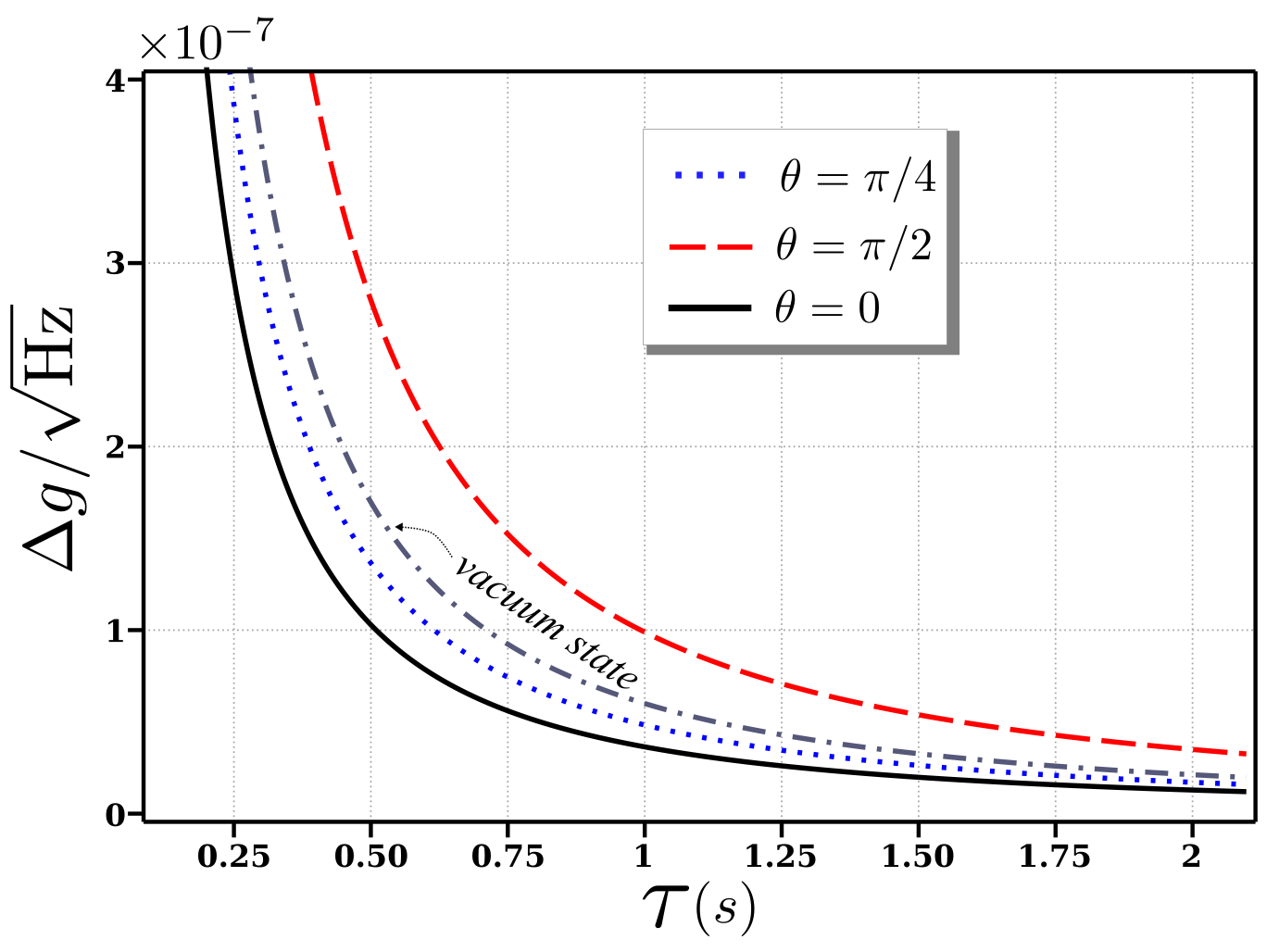}
\caption{Sensitivity for estimating gravitational acceleration $g$ as a function of evolution time $\tau$. The plot compares probes prepared with a fixed squeezing of 4.3~dB ($r=0.5$). The solid black line corresponds to squeezing applied along the canonical position quadrature direction ($\theta = 0$), the dashed red line represents squeezing along the momentum quadrature ($\theta = \pi/2$), and the dotted blue line depicts a probe exhibiting position-momentum correlations, with the squeezing direction chosen at an intermediate phase of $\theta = \pi/4$. The shot-noise limit (indicated by the dot-dashed gray line) is shown for reference. We simulate an ensemble of Cesium atoms with a mass of $m=2.21\times 10^{-25}$~kg, a position uncertainty of $\sigma_0=30$~nm, and for a free-fall time of 500~ms~\cite{Gerginov_2025Metrologia}, yielding a sensitivity of $1.0\times 10^{-7}$~m$\cdot$s$^{-2}/\sqrt{\text{Hz}}$ that is consistent with current technology~\cite{Qvarfort2018}.}
\label{sensitivity}
\end{figure}

To quantify the achievable precision, we use the metrological sensitivity, $\eta = \sqrt{\tau/\mathcal{F}_g}$. This quantity normalizes the estimation uncertainty, $\Delta g \ge 1/\sqrt{\mathcal{F}_g}$, by the square root of the interrogation time, accounting for the statistical improvement from longer measurements (where $\Delta g \propto 1/\sqrt{\tau}$). The result is a time-independent figure of merit that enables a fair comparison between the intrinsic performance of different metrological protocols. Figure~\ref{sensitivity} shows the sensitivity in estimating the gravitational acceleration $g$ as a function of the evolution time $\tau$. The solid black line corresponds to the squeezing applied along the canonical position quadrature direction ($\theta = 0$), the dashed red line represents the squeezing along the momentum quadrature ($\theta = \pi/2$), and the dotted blue line depicts a probe exhibiting position-momentum correlations, with the squeezing direction chosen at an intermediate phase of $\theta = \pi/4$. The dot-dashed gray line indicates the shot-noise limit for comparison. Simulations use parameters suitable for cold Cesium atoms~\cite{Gerginov_2025Metrologia}: a mass of $m=2.21\times 10^{-25}$~kg, position uncertainty $\sigma_0=30$ nm, a squeezing level of 4.3~dB ($r=0.5$), and a free-fall time of 500~ms, which implies a sensitivity of the order of $1.0\times 10^{-7}$~$\text{m}\cdot\text{s}^{-2}/\sqrt{\text{Hz}}$ which is consistent with the performance of current technologies~\cite{Qvarfort2018}.

The relative QFI~(RQFI) is defined as follows: \begin{equation} \mathcal{Q}(\tau,r,\theta)=\frac{\mathcal{F}_{g}(\tau,r,\theta)}{\mathcal{F}_{g}^{\text{vac}}(\tau)}.
\end{equation}
The quantity $\mathcal{Q}$ quantifies the advantage offered by squeezing in the estimation of the gravitational acceleration $g$, relative to a vacuum-state probe. A value of $\mathcal{Q} > 1$ indicates that the use of squeezed states leads to enhanced estimation precision compared to the vacuum state. Conversely, $\mathcal{Q} < 1$ implies that squeezing degrades the estimation performance, whereas $\mathcal{Q} = 1$ indicates that the squeezed and vacuum probes yield equivalent sensitivities. Therefore, we obtain the following:

\begin{gather}
\mathcal{Q}(\tau,r,\theta) =
\frac{[\sinh^{2}(2r)\sin^{2}(2\theta)+1]\tau^4}{[\cosh(2r)-\sinh(2r)\cos(2\theta)](\tau^4+16\tau^2_0\tau^2)} 
+\frac{8\sinh(2r)\sin(2\theta)\tau_0\tau^{3}}{(\tau^4+16\tau^2_0\tau^2)} 
+\frac{16[\cosh(2r)-\sinh(2r)\cos(2\theta)]\tau_{0}^{2}\tau^{2}}{(\tau^4+16\tau^2_0\tau^2)}.
\label{QQ}
\end{gather}

To analyze this expression, we consider its behavior in the short-time limit, explicitly characterized by the condition $\tau \ll \tau_0$:
\begin{gather}
\mathcal{Q}^{\tau\ll\tau_0}(\tau,r,\theta)\approx\frac{\sigma^{2}}{\sigma_{0}^{2}}+\frac{\sinh(2r)\sin(2\theta)}{2\tau_{0}}\tau+\frac{\sinh(2r)\cos(2\theta)}{8\tau_{0}^{2}}\tau^{2}+\frac{\sinh(2r)\sin(2\theta)}{32\tau_{0}^{3}}\tau^{3}+\mathcal{O}(\tau^{4}),
\end{gather}
which, for specific choices of the squeezing phase, namely $\theta = 0$, $\theta = \pi/2$, and $\theta = \pi/4$, simplifies, respectively, to:

\begin{equation}
\mathcal{Q}^{\tau\ll\tau_0}(\tau,r,0)\approx e^{-2r}+\frac{\sinh(2r)\tau^{2}}{8\tau_{0}^{2}},
\end{equation}
\begin{equation}
\mathcal{Q}^{\tau\ll\tau_0}(\tau,r,\pi/2)\approx e^{2r}-\frac{\sinh(2r)\tau^{2}}{8\tau_{0}^{2}},
\end{equation}
\begin{equation}
\mathcal{Q}^{\tau\ll\tau_0}(\tau,r,\pi/4)\approx\cosh(2r)-\frac{\sinh(2r)}{2\tau_{0}^{2}}\tau+\frac{\sinh(2r)}{32\tau_{0}^{3}}\tau^{3}.
\end{equation}
A direct comparison shows that, in the limit $\tau \rightarrow 0$, a squeezing phase of $\theta=\pi/2$ provides a greater metrological advantage for the same squeezing amplitude. On the other hand, in the long-time limit ($\tau \gg \tau_0$), the RQFI becomes

\begin{gather}
\mathcal{Q}^{\tau\gg\tau_0}(\tau,r,\theta) \approx\frac{[\sinh^{2}(2r)\sin^{2}(2\theta)+1]\sigma_{0}^{2}}{\sigma^{2}} 
+\frac{8\sinh(2r)\sin(2\theta)\tau_{0}}{\tau}
-\frac{32\sinh(2r)\cos(2\theta)\tau_{0}^{2}}{\tau^{2}} \nonumber \\ -\frac{128\sinh(2r)\sin(2\theta)\tau_{0}^{3}}{\tau^{3}}+
\mathcal{O}(\frac{1}{\tau^{4}}).
\end{gather}

Again, for $\theta = 0$, $\pi/2$, and $\pi/4$, the expression reduces to:

\begin{equation}
\mathcal{Q}^{\tau\gg\tau_0}(\tau,r,0)\approx e^{2r}-\frac{32\sinh(2r)\tau_{0}^{2}}{\tau},
\end{equation}
\begin{equation}
\mathcal{Q}^{\tau\gg\tau_0}(\tau,r,\pi/2)\approx e^{-2r}+\frac{32\sinh(2r)\tau_{0}^{2}}{\tau},
\end{equation}
\begin{equation}
\mathcal{Q}^{\tau\gg\tau_0}(\tau,r,\pi/4)\approx\cosh^{2}(2r)+\frac{8\sinh(2r)\tau_{0}}{\tau}-\frac{128\sinh(2r)\tau_{0}^{3}}{\tau^{3}}.
\end{equation}

Although the optimal squeeze phase to maximize metrological advantage shifts from $\theta=\pi/2$ in the short-time limit ($\tau \to 0$) to $\theta=0$ in the long-time limit ($\tau \to \infty$), an intermediate phase such as $\theta=\pi/4$ presents a notable trade-off. Although it is never the superior choice at either extreme, it consistently surpasses the shot-noise limit, thereby offering a reliable metrological advantage across the entire time domain.

\section{Measurement schemes}

Although QFI provides the best sensitivity over all POVM measurements, it is crucial to determine which measurement types saturate this constraint \cite{PaternostroARXIV2025}. To achieve this, we take into account local Gaussian projective measurements, which can be calculated simply by updating the covariance matrix as $\widetilde{\boldsymbol{\sigma}} = \boldsymbol{\sigma} + \boldsymbol{\sigma}_m$ \cite{PaternostroPRA2017,PaternostroARXIV2025}, where $\boldsymbol{\sigma}_m = \text{Diag}[s,s^{-1}]/2$ represents the projective measurement and is parametrized by $s$, with measurements of position ($s \rightarrow 0$) and momentum quadratures ($s \rightarrow \infty$), homodyne and balanced heterodyne detection ($s = 1$). For a Gaussian state, the classical Fisher information (CFI) that corresponds to this measuring strategy is \cite{PaternostroPRA2017,PaternostroARXIV2025} 

\begin{equation}
\mathcal{I}_{\alpha}(\widetilde{\boldsymbol{\sigma}},\mathbf{d})=\frac{\text{Tr}[(\widetilde{\boldsymbol{\sigma}}^{-1}\partial_{\alpha}\widetilde{\boldsymbol{\sigma}})^{2}]}{2}+(\partial_{\alpha}\boldsymbol{d})^{\text{T}}(\widetilde{\boldsymbol{\sigma}}^{-1})(\partial_{\alpha}\boldsymbol{d}).
\label{eq:CFIgauss}
\end{equation}
Consequently, the goal of $\mathcal{I}_{\alpha}(\widetilde{\boldsymbol{\sigma}},\mathbf{d})$ is to dynamically examine, by varying the parameter $s$, which measurement approach is optimal to reach the upper limit imposed by quantum mechanics.  For the estimation of gravitational acceleration $g$, considering an initial squeezed vacuum state, the CFI corresponding to different projective measurement schemes $s$ is expressed as:
\begin{equation}
\mathcal{I}_{g}(\tau,r,\theta,s)=\frac{8\tau^{2}s\sigma^{2}m^{4}(2\sigma^{2}+s)+16\hslash\gamma m^{3}s\sigma^{2}+s\tau^{3}+\tau^{3}m^{2}(\hslash^{2}(\gamma^{2}+1)s+2\sigma^{2})}{2\hbar^{2}s^{2}m^{2}(\gamma^{2}+1)+4m^{2}s\sigma^{2}+8m^{2}\sigma^{4}+4m^{2}s\sigma^{2}+8\hslash\gamma m\tau\sigma^{2}+2\tau^{2}\hslash^{2}(\gamma^{2}+1)}.
\end{equation}
Therefore, the CFI for the projective position ($s \to 0$), momentum ($s \to \infty$), and heterodyne ($s \to 1$) measurements, respectively, becomes 
\begin{equation}
\mathcal{I}_{g}^{s\rightarrow0}(\tau,r,\theta)=\frac{\tau^{4}m^{2}\sigma^{2}}{\hslash^{2}\gamma^{2}\tau^{2}+4\hslash\gamma m\sigma^{2}\tau+4m^{2}\sigma^{4}+\tau^{2}\hslash^{2}},
\end{equation}
\begin{equation}
\mathcal{I}_{g}^{s\rightarrow\infty}(\tau,r,\theta)=\frac{4\tau^{2}m^{2}\sigma^{2}}{\hslash^{2}\left(\gamma^{2}+1\right)},    
\end{equation}
\begin{equation}
\mathcal{I}_{g}^{s\rightarrow1}(\tau,r,\theta)=\frac{4\tau^{2}\left(2\sigma^{4}+\sigma^{2}\right)m^{4}+4\tau^{3}\sigma^{2}\hslash\gamma m^{3}+\tau^{4}\left(\frac{\hslash^{2}(\gamma^{2}+1)}{2}+\sigma^{2}\right)m^{2}}{(\gamma^{2}+2\sigma^{2}+1)m\hslash^{2}+4m\sigma^{4}+2m\sigma^{2}+4\hslash\gamma\mathrm{2}m\sigma^{2}\tau+\tau^{2}\hslash^{2}(\gamma^{2}+1)},
\end{equation}
with $\sigma=\sigma_{0}[\cosh(2r)-\sinh(2r)\cos(2\theta)]^{1/2}$ and $\gamma = \sinh(2r)\sin(2\theta)$. To compare which scheme achieves the upper quantum limit provided by the QFI, we introduce the ratio $\mathcal{R}(\tau,r,\theta,s) = \mathcal{I}_{g}(\tau,r,\theta,s)/\mathcal{F}_g(\tau,r,\theta)$, which relates the CFI, $\mathcal{I}_{g}(\tau, r, \theta, s)$, to the QFI $\mathcal{F}_{g}(\tau, r, \theta)$ bound:
\begin{equation}
\mathcal{R}(\tau,r,\theta,s)=\frac{8\tau^{2}m^{2}\sigma^{2}\hslash^{2}\left[\left(\frac{\Gamma\hslash^{2}s}{8}+\frac{\sigma^{2}}{4}\right)\tau^{2}+\gamma\hslash ms\tau\sigma^{2}+s\sigma^{2}\left(2\sigma^{2}+s\right)m^{2}\right]}{\left[8\left(m^{2}(2s\sigma^{2}+s^{2}\Gamma)+\tau^{2}\Gamma\right)\hslash^{2}+32\gamma\hslash m\tau\sigma^{2}+16m^{2}\sigma^{2}\left(2\sigma^{2}+s\right)\right]\left(\frac{\Gamma\tau^{2}\hslash^{2}}{16}+\frac{\gamma\hslash m\tau\sigma^{2}}{2}+m^{2}\sigma^{4}\right)\tau^{2}},
\end{equation}
where $\Gamma=\gamma^{2}+1$. Here, the goal is to determine the conditions and time regimes under which this quantity reaches or approaches unity. As noted in the main manuscript, we find that only the projective momentum measurement ($s \to \infty$) achieves this objective, where the optimal choice for the squeezing direction $\theta$ is time dependent.

\nocite{*}

\end{document}